\documentclass[twocolumn]{svjour3}          
\smartqed  
\usepackage{graphicx}
\usepackage{amsmath}
\usepackage{fixltx2e}
\usepackage[sort]{natbib}
\usepackage{times}
\usepackage{hyperref}

\journalname{Cognitive Neurodynamics}

\begin{document}

\title{Modeling Effect of GABAergic Current in a Basal Ganglia Computational Model.}


\author{Felix Njap
\and Jens Christian Claussen
\and  Andreas Moser \and Ulrich G. Hofmann}
\date{ Published in: Cogn. Neurodyn. (2012) 6:333-341}

\institute{Felix Njap \at
 University of Luebeck\\
	    Graduate School for Computing in Medicine and Life Sciences\\
	    D-23538 L\"ubeck \\
\email{fnajp2001@gmail.com}
	    \and
	    Jens Christian Claussen \at
	    University of Luebeck \\
	    Institute for Neuro- and Bioinformatics\\
	    D-23538 L\"ubeck, Germany \\
	    \email{claussen@inb.uni-luebeck.de}
	    \and	
	    Andreas Moser \at
Department of Neurology, University of L\"ubeck,
\\	    D-23538 L\"ubeck, Germany \\
	    \email{andreas.moser@neuro.uni-luebeck.de} 
\and
Ulrich G.\ Hofmann
\at 
Institute for Signal Processing,  University of Luebeck \\
23538 L\"ubeck, Germany \\
\email{hofmann@isip.uni-luebeck.de}
}
\date{Received: 21.\ July 2011 / Revised: 13 March 2012 /  Accepted: 16
April 2012\\
Published in: Cognitive Neurodynamics (Springer) 
\href{http://www.springerlink.com/content/1871-4080/}{\bf Cognitive
 Neurodynamics, Volume 6, 633 (2012)} 
}

\maketitle


\begin{abstract}
Electrical high frequency stimulation (HFS) of deep brain regions is a method shown to be clinically effective in different types of movement and neurological disorders. In order to shed light on its mode of action a computational model of the basal ganglia network coupled the HFS as injection current into the cells of the subthalamic nucleus (STN). Its overall increased activity rendered a faithful transmission of sensorimotor input through thalamo-cortical relay cells possible. Our contribution uses this model by Rubin and Terman [\textit{J. Comput. Neurosci.,} \textbf{16}, 211-223 (2004)] as a starting point and integrates recent findings on the importance of the extracellular concentrations of the inhibiting neurotransmitter GABA. We are able to show in this computational study that besides electrical stimulation a high concentration of GABA and its resulting conductivity in STN cells is able to re-establish faithful thalamocortical relaying, which otherwise broke down in the simulated parkinsonian state.
\end{abstract}
\keywords{Computational model\and  Synaptic conductances \and $\gamma$-Aminobutric acid \and Deep bbrain stimulation \and Parkinsonian condition}

\section{Introduction}
Deep brain stimulation (DBS), especially of the subthalamic nucleus (STN), utilizing chronically implanted electrodes has become an effective, though symptomatic, therapy for a wide range of neurological disorders \cite{benabid02, benabid07, mc04, deep01}. However the detailed working mechanism regarding molecular and pharmacological aspects is not yet fully understood. During the past decade computational neuroscience has attempted to shed light on the mechanism of DBS by numerical simulations to optimize the therapeutic outcome of DBS in movement disorders \cite{alejandro06,  Pirini09}. With this aim a cellular-based model of the basal ganglia system was implemented by \cite{rub04}. The original model was able to reproduce the physiological and pathological activities of STN and thalamus cells (TC) in a realistic basal ganglia network and proposed such an explanation for the reduction of parkinsonian symptoms under electrical HFS. Their findings are based on increasing the firing activity of STN rather than shutting it down. \\

The following study utilizes the original model, but develops it further in light of recently presented neurochemical findings on the DBS rationale. Those experiments quantitatively measured extracellular neurotransmitter concentrations, and showed that electrical high frequency stimulation (HFS) induced selective γ-aminobutyric acid (GABA) release as a mechanistic basis of HFS \cite{man06, man09, hiller07, feuerstein11}. Whereas usually DBS is considered to provide excitatory input to STN neurons leading to an increased activity, we replaced it with inhibitory postsynaptic current (IPSC) exclusively conveyed by GABA \cite{gerstner02, hutt10, forster08, liu10}.  Our current study seeks to numerically examine the thalamus' output response under DBS-related current and compare this to output response with GABAergic currents applied to the same target cells in STN instead.

\section{Method}
Our model follows up on the seminal model of the basal ganglia thalamic network by Rubin and Terman and the modified version described in our recent contribution. Each cell type in our model network is described by a single compartment and has Hodgkin-Huxley-type spike generating currents as described previously \cite{njap11}. A detailed description including all parameters and nonlinear equations has been published elsewhere (e.g., \cite{guo08}).
The voltage in the original Rubin and Terman's original model obeys the following equation (\ref{eq:DBS}):
\begin{equation}
\label{eq:DBS}
C_{m}\frac{dV}{dt}=- I_{Na}-I_{K}-I_{Ca}-I_{T}-I_{AHP}-I_{Leak}-I_{Syn}+I_{DBS},
\end{equation}
The model features: potassium and sodium spike-producing currents $I_{K}$, $I_{Na}$; a low-threshold T-type $\left( Ca^{2+}\right)$ current $\left(I_{T}\right)$; a high-threshold $\left( Ca^{2+}\right) $ current $\left(I_{Ca}\right) $; a $\left( Ca^{2+}\right)$ activated, voltage-independent after hyper polarization $\left( K^{+}\right) $ current $\left(I_{AHP}\right)$, and a leak current $\left(I_{Leak}\right)$. All these currents are described by Hodgkin-Huxley formalism. $I_{DBS}$ represents the deep brain stimulation current of the STN modeled with the following periodic step function (\ref{eq:IDBS}):
\begin{equation}
\label{eq:IDBS}
I_{DBS}=i_{D}\left( \sin\left( 2\pi t/\rho_{D}\right) \right) \left( 1-\theta \left( \sin \left( t+\delta_{D}\right) /\rho_{D}\right) \right),
\end{equation}
where $ i_{D}$ is the stimulation amplitude, $\rho_{D}$ stimulation period, $ \delta_{D}$ duration of each impulse, and $\theta$ represents the Heaviside step function given by (\ref{eq:heav}):
\begin{equation}
\label{eq:heav}
\Theta(x)= \left \{ \begin{array}{lll}
0 & {\ \ \rm for \ \ } x<0 \\
&\\
\frac{1}{2} & {\ \ \rm for \ \ } x=0,\\
&\\
1 & {\ \ \rm for \ \ } x>0 \\
\end{array}
\right.
\end{equation}
In the original model, during stimulation $ i_{DBS}\left(t\right)$ was taken as a large positive constant and was applied directly to the neuronal membrane in the STN neuron model. In our current study the new membrane potential of each STN neuron integrates over additional ion channels and stochastic $ C_{m}\eta$ to obtain more realistic simulations but also to account for a specific type of experimentally recorded pattern which can not be seen in purely deterministic simulations \cite{braun00}.  It obeys the following equation  (\ref{eq:STN}):
\begin{equation}
\label{eq:STN}
C_{m}\frac{dV}{dt}=- I_{Na}-I_{K}-I_{Ca}-I_{T}-I_{AHP}-I_{Leak}-I_{Syn}+C_{m}\eta
\end{equation}
where $I_{syn}\left( t\right)=g_{syn}\left(t\right) s\left( V-E_{syn}\right)$ represents an inhibitory channels where $s$ ￼satisfies the stochastic differential equation \cite{higham01} (\ref{eq:sto}):
\begin{equation}
\label{eq:sto}
ds=\left[\alpha \left( 1-s\right) -\beta ss \right] dt +\sigma d\eta ,
\end{equation}
$\alpha$ and $\beta$ are the forward and backward rate constants, and the Gaussian white noise included is characterized by mean $<\eta =0>$, and variances $<\eta\left(t \right) \eta \left( 0\right) >=2\sigma \delta \left( t\right)$, with $\sigma$ the strength of the noise. The noise strength $\sigma$ decreases as the square root of the number of ions channels \cite{fox97} (\ref{eq:noise}):
\begin{equation}
\label{eq:noise}
\sigma =\left[ \alpha \left( 1-s\right) +\beta s\right]/\left[ \left( \tau N_{s}\right)\right]^{1/2} 
\end{equation}
with parameters $\alpha = 1\:msec^{-1}$, $\tau =100\:msec$, $N_{s}=500$ 
and is chosen such that $\alpha/\left( \alpha+\beta\right) =0.2$, 
$\sigma$ is expressed in units 
of $mV^{2}/msec$, time in msec, currents in $\mu A/cm^{2}$. 
In all our simulations parameter values were chosen to produce 
reasonable visual agreement with experimental records. 

\begin{figure}[htpb]
  \centering
    \centerline{\includegraphics[width=09.55cm]{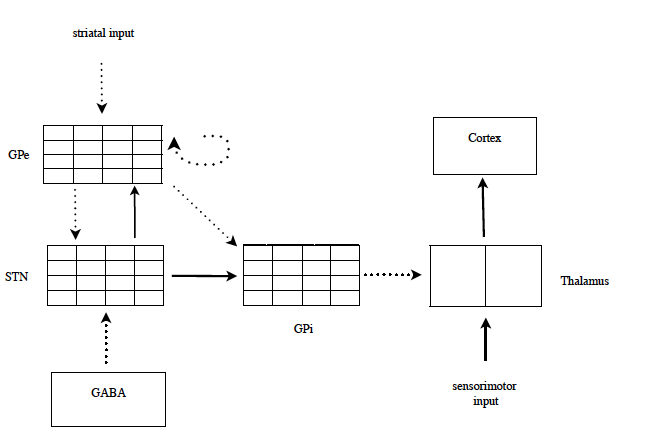}}
\caption[Network model consisting of sixteen STN, GPe and GPi and two TC]{Network model consisting of sixteen STN, GPe and GPi and two TC. The thalamus receives synaptic inhibition from GPi and excitatory input, related to sensorimotor activity. GPi and GPe both receive excitatory input from STN, and GPi receives inhibition from GPe. There is interpallidal inhibition among GPe neurons. STN receives inhibition from GPe and GABAergic currents.  Each STN neuron receives inhibitory input from two GPe neurons. Each GPe neuron receives excitatory input from three STN neurons and inhibitory input from two other GPe neurons. Each GPi neuron receives excitatory input from one STN neuron. Each thalamic neuron receives inhibitory input from eight GPi neurons. The thalamus is viewed as a \textit{relay station} where cells have the unique role of responding faithfully to each excitatory sensorimotor input. GPe receives striatal input. Adapted from \cite{rub04}.}
\label{fig:rub}
\end{figure}

The parameter $E_{syn}$ and the function $g_{syn}\left( t\right) $ can be used to characterize different types of synapses. The parameters that describe the conductivity of transmitter-activated ion channels at a certain synapse are chosen in such a way as to mimic a time course and the amplitude of experimentally observed spontaneous postsynaptic currents. In this paper, to take heterogeneity into account the conductance current is described by the following kinetic equation (\ref{eq:syn}):
\begin{equation}
\label{eq:syn}
g_{syn}\left( t\right) =\sum\limits_{f}\bar{g}_{{syn}}e^{-\left( t-t^{\left( f\right) }\right)\tau}\:\theta \left( t-t^{\left( f\right) }\right),
\end{equation}
where  $t^{\left(f \right) }$ denotes the arrival time of a presynaptic action potential which follow a Poisson distribution at rate 0.05 spikes/sec. $E_{syn}=-75\:mV$ is the reversal potential, $\tau = 5\:ms$ is the time constant and $\bar{g}_{{syn}}$ is the amplitude describing the maximal synaptic conductance of GABA. The time-dependent conductance of inhibitory synapses in deep cerebellar nuclei can be described by a simple exponential decay given by Eqn. (\ref{eq:syn}).\\
In the basal ganglia, the majority of neurons uses GABA as neurotransmitter and has inhibitory effects on their targets \cite{chak10,  boyes07}. For the sake of computation and simplicity, our model included only the slow component $GABA_{A}$ synapse. Parameter values were the same as in Rubin and Terman's model. All simulations were performed using the software XPPAUT written by G. Bard Ermentrout \cite{ermentrout02} and MATLAB for analysis. The numerical method used was an adaptive-step fourth order Runge-Kutta method (\textit{Qualst.RK4 in XPP}) with a typical time step of 0.01 msec. \\
The model network is depicted in Fig. \ref{fig:rub}. It consists of five anatomical nuclei representing the external segment of globus pallidus (GPe), subthalamic nucleus (STN), the internal segment of globus pallidus (GPi), thalamus, and cortex, where the first three nuclei belong to the basal ganglia network. Arrows with dashed lines indicate inhibitory synaptic connections and inputs, whereas solid lines indicate excitatory synaptic connection and inputs. In the original model GPe and GPi neurons were biased with applied currents that varied between the normal-healthy and parkinsonian conditions, modeling changes in the strength of striatal inhibition.

To simulate a parkinsonian state, parameters were chosen to reproduce the 
behavior of experimentally recorded 
cells ~ of ~ an ~
1-methyl-4-phenyl-1,2,3,6-tetrahydropyridine
\\
(MPTP), non-human primate model of Parkinson’ diseases (PD) shown in \cite{guo08}. We then applied GABA-mediated currents onto STN cells during this parkinsonian condition and observed the direct effect on thalamus cells (TC) to relay sensorimotor input to the cortex. Secondly, as Rubin and Terman did, we observed the thalamus' ability to faithfully relay sensorimotor input to the cortex as we slowly increased the synaptic conductance of GABA. Our model's output was compared to the optimal DBS current described in the original model. Thus we were finally able to evaluate our network model using two performance scores (\textit{error index} and \textit{coefficient of variation}) with the aim of measuring thalamocortical cell responsiveness to stimulus input.
\section{Results}
\subsection{Normal and Parkinsonian firing patterns}
\begin{figure}[htpb]
  \centering
    \centerline{\includegraphics[width=09.55cm]{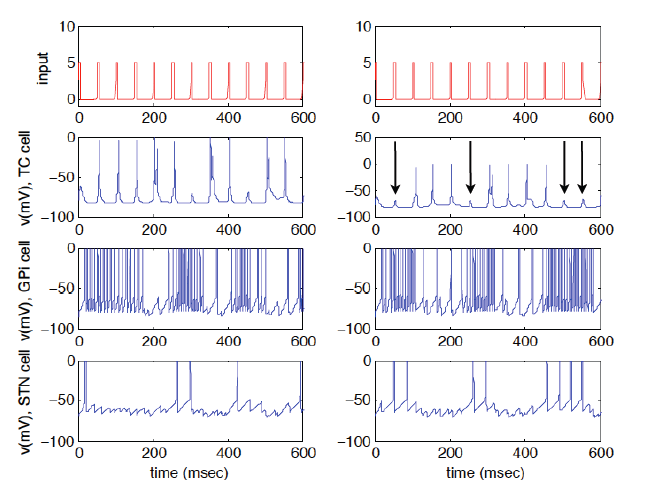}}
 \caption[Periodic sensorimotor stimulation input and TC responsiveness in the two clinical states]{Periodic sensorimotor stimulation input and TC responsiveness in the two clinical states: normal (left) and parkinsonian (right) and corresponding GPi and STN activity under these conditions. Arrows indicate dropped responses of TC on driving input, thus indicating poor information transfer to the cortex. During the normal state, STN neurons fire irregularly whereas in parkinsonian state, each STN neuron fires in a periodic tremor-like fashion, thus leading STN neurons’ populations to break up into two clusters. Adapted from Adapted from \cite{rub04}.}
  \label{fig:norpark}
\end{figure}

Although the network parameters are set to produce the parkinsonian state in the absence of DBS, Rubin’s DBS model show that the presence of electrical DBS restores the faithful relay of inputs to the cortex by the TC. On the other hand, during parkinsonian condition, the thalamus is no longer able to relay sensorimotor input faithfully due to the bursting activity of GPi. This tonical activity may considerably influence thalamic responsiveness activity. At this stage we replaced DBS current with inhibitory postsynaptic current (IPSC) in STN cells and questioned the ability of thalamus to produce the same output. Fig. \ref{fig:dbsgaba}. shows that the thalamus cells produce similar network effects when replacing DBS excitatory input with the GABAergic inhibitory current at higher synaptic conductances. The loss of connectivity observed in parkinsonian state Fig. \ref{fig:norpark}  is restored, corroborating the key role of synaptic inhibition. 

\subsection{DBS acts excitatory and $GABA_{A}$-type currents inhibitory}
\begin{figure}[htpb]
 \centering
    \centerline{\includegraphics[width=09.55cm]{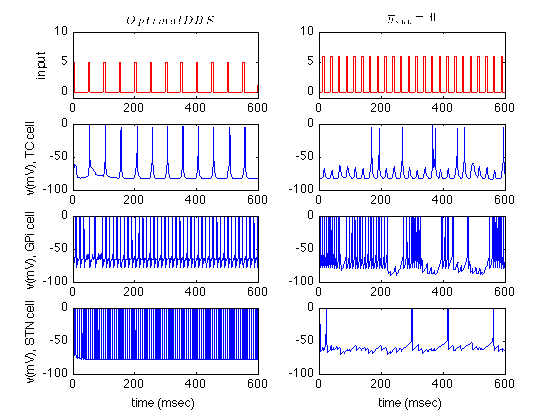}}
     \caption[ Periodic sensorimotor stimulation input and TC cells responsiveness in the optimal DBS stimulation (left) and with vanishing GABA   ]{Periodic sensorimotor stimulation input and TC cells responsiveness in the optimal DBS stimulation (left) and with vanishing GABA synaptic conductance amplitude (right). Electrical HFS increases activity in the STN \cite{garcia05}, thus leading to increased tonic activity in GPi cells. This seemingly contradicts the idea that electrical HFS is a way of silencing the pathologically overactive indirect pathway as it is done in therapeutic lessoning \cite{olanow00}. Under stimulation conditions, DBS restores the thalamus ability to transmit information, whereas with vanishing synaptic conductance amplitude, TC cells are unable to transmit information to the cortex.}
     \label{fig:dbsgaba}
\end{figure}
\begin{figure}[htpb]
 \centering
    \centerline{\includegraphics[width=09.55cm]{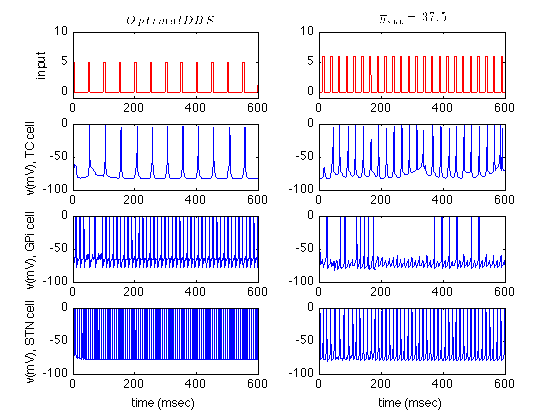}}
     \caption[The output model membrane potential of TC, GPi and STN cells with optimal DBS stimulation]{The output model membrane potential of TC, GPi and STN cells with optimal DBS stimulation (left) and GABAergic tonic inhibition (right).}
      \label{fig:dbsgaba37}
   \end{figure}

Thalamus cells are not able to relay information to the cortex as seen in Fig. \ref{fig:dbsgaba}. (right) with vanishing synaptic $GABA_{A}=0$ currents. Therefore, we limited our simulations in the beginning to two different input regimes consisting of smaller and larger synaptic inputs. Increasing the synaptic conductance up to 40 pS, our simulations results show, that TC cell relay fidelity is qualitatively restored the same way as DBS is able to do Fig. \ref{fig:dbsgaba37}. Above $\bar{g}_{{syn}}=40$ our networks are no longer stable. The loss of faithful relaying quickly returns for small values of synaptic conductance see Fig. \ref{fig:gaba102030}. 
  In order to quantify TC cells' output in term of correct responses, we utilized two performance scores: \textit{The error index (EI)} and the \textit{coefficient of variation (CV)}. 
 As Rubin and Terman defined the $EI$ is the total number of errors divided by the total number of input stimuli. The $CV$ is the ratio of standard deviation to the average of the inter-spike intervals. 
 
 This coefficient provides information on the output of the thalamocortical cells. By thresholding ($V_{th}=-45\:mV$) the membrane potential, we defined the thalamic cell spike response to each stimulus amplitude. For a perfect relay of our periodic input we expect $CV=0$ and $EI=0$. This ideal scenario corresponds to constant inter-spike intervals, but $EI=0$ is not incompatible with $CV\neq0$ (for example if pulses are relayed but timing is slightly affected). On the other hand, an $EI\neq 0$ and $CV=0$ correspond to one stimulus pulse relayed periodically every two or more input pulses \cite{alejandro06}. Table \ref{tab:table41}. lists different values of $CV$, and $EI$ in normal state, parkinsonian state, optimal applied DBS current and with the GABA input that seems to produce the most beneficial therapeutic effect in our simulation. 
  \begin{table}[ht]
  \caption{Averaged error index and coefficient of variation.}
  \label{tab:table41}
  \centering
  \begin{tabular}{c c c c c}  \hline \hline
                 & \textbf{Normal}  &\textbf{Parkinson} & \textbf{Optimal DBS}& \textbf{$\bar{g}_{{syn}}=37.5$}\\\hline
    $EI$      & 0.0612       & 0.6265 & 0.1734  &0.0949 \\                      
   $CV $ &0.7318  &0.6602& 0.3087&0.3048\\
  \hline
   \end{tabular}
\end{table}
 To analyze the results displayed in Fig. \ref{fig:dbsgaba37}, we successively introduced the coherence measure taken from \cite{white98} within and between spike trains of basal ganglia different nuclei when synaptic conductances fluctuate and computed the coherence reduction  (CR) described by \cite{moran12}. A measure of coherence usually characterized the functional integration between the different components of the brain. The correlation or the coherence measure determines the level of synchrony and quantifies the linear correlation in time-frequency domain. Therefore, one distinguishes the magnitude square coherence function and the phase function. Spike trains were approximated by a series of square pulses of unit height and fixed width of $20\%$ of the period of the most rapid firing cell. Each square wave is centered around the peak of the individual action potentials in the train. Then, we computed the shared area of the square pulses from each train that partly coincide in time. The cross-correlation at zero time lag was considered. This correspond to the evaluation of the shared area of the unit-height pulses. Finally, we took the coherence as the sum of these shared areas, divided by the square root of the product of the summed areas of each individual pulse train \cite{baker02}. 
 
 If $x\left(t \right)$ is the series of unit height pulses for the first cell over N time steps and $y\left(t \right)$ is the series of pulses for the second cell, then the coherence measure or correlation in time-domain is calculated as  (\ref{eq:coh}):
 \begin{equation}
\label{eq:coh}
Coherence\: Measure=\frac{\sum\limits_{i=1}^{N}x\left(t \right)\ast y\left(t \right)}{\sqrt{\sum\limits_{i=1}^{N}x\left(t \right)\ast \sqrt{\sum\limits_{i=1}^{N}y\left(t \right)}}},
\end{equation}

Given the spike-trains $x\left(t \right)$ and $y\left(t \right)$, their Fourier transforms $X\left(\omega \right)$ and $Y\left(\omega \right)$, and complex conjugates $X^{\ast}\left(\omega \right)$ and $Y^{\ast}\left(\omega \right)$, the coherence is readily computed in the frequency domain as (\ref{eq:cohf}):
 \begin{equation}
\label{eq:cohf}
C_{XY}\left( \omega\right)=\frac{C_{XY}\left( \omega\right)}{\sqrt{P_{XX}\left( \omega\right)P_{YY}\left( \omega\right)}},
\end{equation}
where $P_{XX}\left( \omega\right)=X\left(\omega \right)X^{\ast}\left(\omega \right)$ is the power spectrum of $x\left(t \right)$, and $P_{XY}\left( \omega\right)=X\left(\omega \right)Y^{\ast}\left(\omega \right)$ is the cross-spectrum of $x\left(t \right)$ and $y\left(t \right)$. In this study, since both models produces spike output, $P_{XX}\left( \omega\right)$, $P_{YY}\left( \omega\right)$ and $P_{XY}\left( \omega\right)$ were computed for the mean spike trains using multi-taper estimation methods on 256 ms windows and discrete sequences every 64 ms over the trial time span. Coherences were then computed from trial-averaged spectra. We note that coherence is only meaningful at frequencies with non-vanishing power.

In all our simulation results, the spiking activity of the GPi cells reduces drastically as the synaptic conductances strength increases. The mean firing rate recorded during deep brain stimulation was 161 spikes per second, whereas with the GABA input producing the most beneficial therapeutic outcomes the mean firing rate was 40 spikes per second. Furthermore, the coherence reduction (CR) in Fig. 4.7. is in line with recently findings by \cite{moran12, wilson11} and supports this hypothesis on declining coherence in neuronal spiking activity within and between different nuclei of the basal ganglia during STN macro-stimulation. Our results are similar to those of \cite{moran12}, but our explanations differ considerably. They saw a decline in coherence during stimulation. They mainly attributed the decline to the STN oscillatory decoupling from the GPi. In our work we controlled for this synaptic coupling factor, using a smaller parameter value less than  $5\%$ differences in afferent synaptic currents $g_{STN\rightarrow GPi}$ rather than in large currents (base value is $0.3 mS/cm^{2}$). This achieved similar relative differences in intrinsic firing rates, and we still saw a drop-off in coherence. Besides, when the synaptic coupling is extremely fast, the coupling frequently cause  neurons towards anti-synchrony \cite{skinner94, wang92}.

\subsection{Effects of different conductances}
\begin{figure}[htpb]
 \centering
    \centerline{\includegraphics[width=09.55cm]{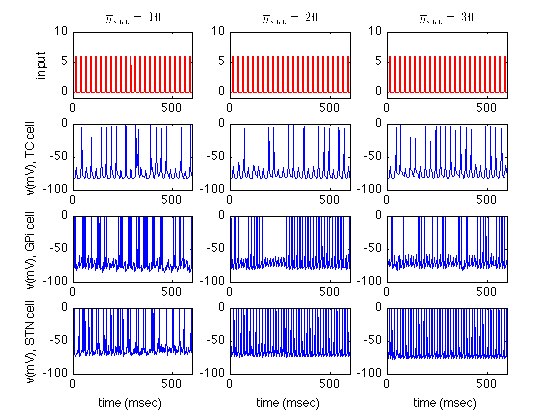}}
     \caption[Cellular activity exhibited by the TC, STN and GPi cell models]{Cellular activity exhibited by the TC, STN and GPi cell models for three different synaptic conductance levels. With an increase in synaptic conductance, the relay properties of TC cells improve as can be seen on the EI and CV, respectively. $EI,\:CV(10)= 0.4241,\:0.4851;\: EI,\:CV(20)= 0.3608,\: 0.5683;\: EI,\:CV(30)= 0.2089,\: 0.5333$.}
     \label{fig:gaba102030}
   \end{figure}
\begin{figure}[htpb]
 \centering
    \centerline{\includegraphics[width=09.55cm]{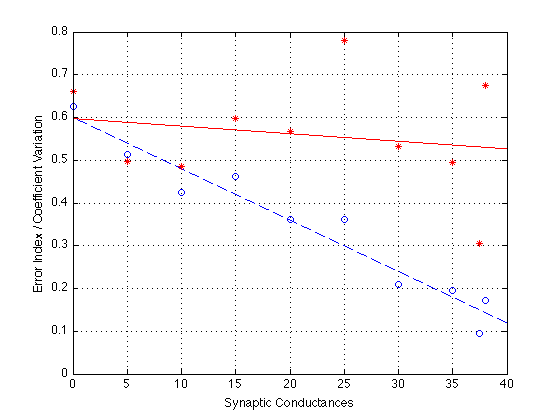}}
     \caption[Coefficient variation and error index]{Shows average error index EI (circles) of the two thalamic cells and (CV) coefficient  of variation (star) dependence of  synaptic conductance \cite{mayer06}. Dashed and solid lines consist of linear interpolation representing lines $\left( y=0.6007-0.0120\ast x\right)$ and $\left( y=0.6007-0.0120\ast x\right) $ best fit which best approximates EI and CV  respectively.}
      \label{fig:error}
   \end{figure}

Fig. \ref{fig:gaba102030} illustrates the model cell dynamics in dependence on synaptic conductance. One observes different effects on the relay properties of TC cells. For an increase in synaptic conductance, TC cells relay properties improve. Fig.  \ref{fig:error} illustrates the effects of the synaptic conductance on the relay properties of TC cells as quantified by the error index (circles) and the coefficient of variation (stars). When taking the average of both TC cell outputs, we found that the EI decreased with increasing inhibition, resulting from a decreasing number of incorrectly transmitted responses. The coefficient of variation does not show a similar tendency to decrease with synaptic connectivity; however similarities with the error index cannot be taken much further.

   The mean frequency-domain coherogram in Fig. 4.7. over 50 trials is 0.4049 with $\bar{g}_{{syn}}=37.5$, whereas the average coherence is 0.9921 when $\bar{g}_{{syn}}=0$. The model thus exhibits a decreased coherence as the synaptic weight parameter increases.  Our numerical simulations show that at gamma frequency 30-80Hz, a significant synchrony is observed, however, with heterogeneous cells, synchrony may not be possible at all frequencies. In particular, a network of this kind seems unlikely to support synchronous firing at a frequency greater than 200Hz (Fig. 4.7.), a frequency to fast to be synchronized by GABAA. The advantage of this approach is that an understanding of the complexity of the nonlinear, interacting dynamics has been obtained using previous theoretical insights on inhibitory network dynamics.
   
   \begin{figure}[htpb]
 \centering
    \centerline{\includegraphics[width=09.55cm]{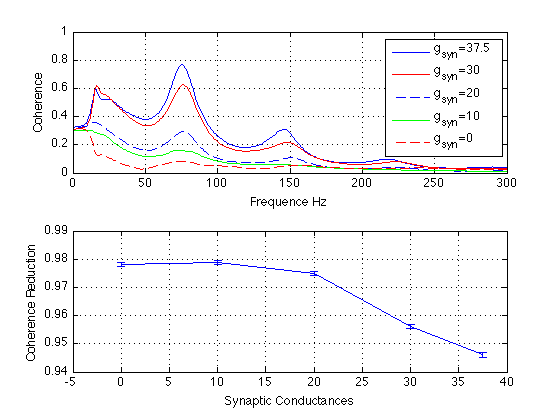}}
     \caption[Mean coherence and coherence reduction]{Mean coherence measure in frequency domain for GPi-STN cells (top panel) and the coherence reduction (CR) over the synapic conductances input (bottom panel).}
     \label{fig:cv}
   \end{figure}
   
\begin{figure}[htpb]
 \centering
    \centerline{\includegraphics[width=09.55cm]{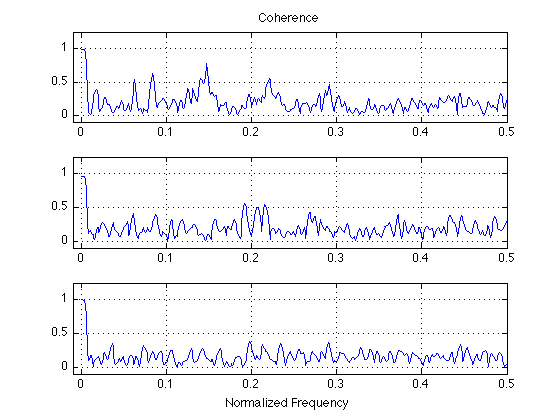}}
     \caption[Coherence for GPi-GPi cells]{Coherence for GPi-GPi cells with the same properties but different synaptic weights as quantified by $\bar{g}_{syn}$=10 (top panel), 20 (middle panel) and 30.}
     \label{fig:coh}
   \end{figure}

Fig. \ref{fig:coh} 
shows the mean coherence within the same target nucleus of the GPi cell. 
Significant coherence is found at a normalized frequency of 0.25 when shifting zero-frequency component to center of spectrum whereas between STN-GPi, it is found at 80 Hz. The response of the network depends on the firing frequency and the time constant of the synaptic weight. Coherence can be reduced in two qualitatively different ways depending on the parameters-either by increasing gradually the synaptic weight coupling parameter, or through suppression, the latter neurons with higher intrinsic rates fire near synchrony and keep their slower counterparts from firing.

   \section{Discussions and Conclusion}
 Recent evidence in HFS research points towards a therapeutic mechanism based on effects on the network activity due to a selective GABA release by electrical HFS \cite{man06, man09, hiller07, feuerstein11}. In this study, we used a computational model of the relevant neural structures to examine the effects of low and high conductance inputs in STN target cells to alleviate tentative symptoms by regularizing the pathological synaptic activity of the basal ganglia output structure, the globus pallidus internus (GPi). We used the averaged error index of the thalamic neurons as surrogate for symptom severity and found that synaptic conductance values below $\bar{g}_{{syn}}=30$ did not regularize GPi synaptic activity, thus did not improve thalamic relay fidelity sufficiently. In contrast, values above $\bar{g}_{{syn}}=30$ did regularize GPi activity, thus significantly improved TC neurons’ relay fidelity.

Synchronized neural activity plays a major role in coding and reliable information transmission \cite{gong10}. Synchronization, however, can be enhanced depending on synaptic network connectivity \cite{qu11} in print, with the extreme case of pathological fully synchronized network activity \cite{Pirini09}. Our study may be useful in studying the intermittent synchrony generated by moderately increased coupling strength in the basal ganglia due to the lack of dopamine and investigate the boundary region between synchronized and nonsynchronized states in PD \cite{park11}.

We conclude that indirect inhibition of neuronal output by means of activation of axon terminals makes the synaptic connectivity with neurons near the stimulating electrode a possible explanation of therapeutic mechanisms of actions of electrical HFS \cite{dostrovsky00, magarinos02}.


\begin{acknowledgements}
This work was supported by the 
``Graduate School for Computing in Medicine and Life Sciences''
funded by Germany's Excellence Initiative [DFG GSC 235/1].
\end{acknowledgements}

\bibliographystyle{plain}
\bibliography{njap_cody}

\end{document}